\documentclass[twocolumn,nofootinbib,aps,showpacs,showkeys]{revtex4}

\usepackage{graphics}
\usepackage{amsfonts}
\usepackage{latexsym}

\begin{document}

\title{``Expansion" around the vacuum equation of state - \\ sudden future singularities and asymptotic behavior}

\author{H. \v Stefan\v ci\'c}
\email{shrvoje@thphys.irb.hr}

\affiliation{Theoretical Physics Division,
Rudjer Bo\v{s}kovi\'{c} Institute, \\
   P.O.Box 180, HR-10002 Zagreb, Croatia}


\begin{abstract}
The dark energy model with the equation of state $p_{d}=-\rho_{d} - A \rho_{d}^{\alpha}$ is studied. The model comprises and provides realization of several types of singularities in different parameter regimes:
the divergence of the dark energy density and pressure at finite time and finite value of the scale factor, the singularity of the ``big rip" type and the sudden future singularity recently introduced by Barrow. For parameter choices which lead to a nonsingular expansion of the universe, various types of the asymptotic evolution are found. The entire time evolution of the universe is described both analytically and numerically. The advantages of this dark energy EOS as a parametrization of dark energy are discussed.
\end{abstract}


\noindent

\maketitle

The diverse cosmological observations, including the supernovae of the type
Ia (SNIa) \cite{SNIa}, the anisotropies of cosmic microwave background
radiation (CMB) \cite{CMB}, and the large scale structure (LSS) \cite{LSS},
among others, indicate that the universe is presently in the state of
accelerated expansion. Unraveling the source of this acceleration has become
one of the highest priorities of theoretical cosmology.
An interesting hypothesis is the assumption of modification of the gravitational
interaction at cosmological scales \cite{modgrav}. However,
a large majority of cosmological models explain the acceleration of the
universe in terms of a component with the negative pressure, the so
called {\em dark energy} \cite{rev}.

The nature of dark energy has not yet been established. Apart from the
negativity of its pressure, very few of its properties have been unambiguously
fixed by the observational data. It is customary to model dark energy as an
ideal homogeneous fluid with an equation of state (EOS)
\begin{equation}
\label{eq:eos}
p_{d} = w \rho_{d} \, ,
\end{equation}
where $\rho_{d}$ and $p_{d}$ represent the energy density and the pressure of
dark energy, respectively. The parameter of the equation of state $w$ is
generally a function of cosmologically relevant quantities.
No matter if the parameter of the EOS is considered constant or is modeled as the dynamic quantity,
the analyses of the cosmological data constrain the present value of the parameter of the dark
energy EOS to an interval around the value $w=-1$ \cite{obser}.


A natural benchmark for the dark energy EOS parameter $w$ is the value
characteristic of the cosmological constant (CC), i.e., $w=-1$.
This benchmark value is very useful for natural classification of
dark energy models. Namely, the dark energy density in the models with $w>-1$ decreases with
the expansion of the universe. For the models with $w=-1$ (CC), the dark
energy density is constant, whereas it grows with the expansion of the universe for
the models with $w < -1$. The cosmological
constant is one of the most popular explanations of the nature of dark
energy \cite{cc}. A huge discrepancy between the value inferred from
the observations and the value predicted by quantum field theory requires
the existence of some unknown fine-tuning mechanisms. The coincidence of the CC
energy density and the energy density of nonrelativistic matter at the
present epoch of the expansion of the universe requires further fine-tuning.
Some of the problems related to the CC may be alleviated in the formalism of the
renormalization group equation for the CC
\cite{Sola,mi,ShSoSt,ReuterGen,ReuterAstro}. This approach is also promising in
the treatment of some astrophysical problems \cite{ShSoSt,ReuterAstro}.
A large class of dark energy models satisfy $w \ge -1$. The prominent members of
this class are {\em quintessence} \cite{Q}, {\em k-essence} \cite{k}, and
{\em Chaplygin gas} \cite{Chaplygin} which incorporates a very attractive
possibility of unifying dark energy and dark matter.
The models satisfying $w \le -1$ are
called {\em phantom energy} models \cite{Caldwell, phantom, phantomja}. Some phantom energy
models include a very interesting ``big rip" event with a dramatic influence on
the destiny of all bound structures \cite{bigrip}. Certain attempts of the microscopic formulation of phantom energy models exhibit the quantum instability of vacuum
\cite{instab}, which further increases the controversy around the concept of phantom energy. 
Some recent considerations also study the possibility that the expansional effects, usually attributed to 
phantom energy, might be mimicked by the more complex dynamics of nonphantom components or 
configurations \cite{phantommimic}. Finally, a legitimate question is whether the parameter of the 
EOS may cross the CC boundary $w=-1$. A recent analysis \cite{vikman} shows that in the framework 
of k-essence dark energy, such a transition is unlikely. It should also be mentioned that interesting 
alternatives explaining superaccelerated expansion of the universe in terms of quantum effects of scalar fields 
in the de Sitter space, requiring no phantom energy, also exist \cite{onemli}.


All intervals for $w$ obtained as the constraints from the observational data
include the $w=-1$ value and significantly overlap with both $w < -1$ and $w>-1$
segments of the $w$ axis.
From the theoretical side, at $w=-1$ there is a
qualitative change in the evolution, as well as in other properties
of dark energy. Therefore, since from both theoretical and observational
perspective the dark energy EOS is ``centered" at the vacuum EOS, it is
useful and interesting to consider dark energy models with the EOS which is also
centered at the vacuum EOS $w=-1$. In this paper we consider a class of such
models where the dark energy EOS is reminiscent of an expansion around the
vacuum EOS. Namely, we consider dark energy models (already
mentioned in \cite{Odin}) with the following equation of state:
\begin{equation}
\label{eq:oureos}
p_{d} = -\rho_{d} - A \rho_{d}^{\alpha} \, .
\end{equation}
Here $A$ and $\alpha$ are real parameters. In our considerations we focus
on the models with
$\alpha \neq 1$ because in the $\alpha=1$ case the dark energy EOS reduces to the dark
energy with the constant EOS of either the quintessence ($A<0$) or the phantom ($A>0$)
type.

Many interesting dark energy models are motivated by the considerations from the string theory \cite{dine},
one specific example being the tachyonic dark energy \cite{sen}. Consideration of the contribution of the
transplanckian degrees of freedom in the framework of string theory \cite{frampton} implies the
possibility of the supernegative equation of state for the dark energy and also further motivates
the analysis of the more general dark energy equation of state (\ref{eq:oureos}).

From the conservation of the energy-momentum tensor of dark energy with the EOS
(\ref{eq:oureos})
\begin{equation}
\label{eq:scalgen}
d \rho_{d} + 3(\rho_{d}+p_{d}) \frac{d \, a}{a} = 0 \, ,
\end{equation}
we obtain the following expression for the scaling of the dark energy density:
\begin{equation}
\label{eq:scalour}
\rho_{d}=\rho_{d,0} \left( 1 + 3 \tilde{A} (1-\alpha) \ln \frac{a}{a_{0}} \right)^{1/(1-\alpha)} \, .
\end{equation}
Here we have introduced an abbreviation $\tilde{A} = A \rho_{d,0}^{\alpha-1}$.
The parameter $\tilde{A}$ determines the type of the behavior of the dark energy density with the expansion of the universe. The speed of change of the dark energy density
\begin{equation}
\label{eq:speed}
\frac{d \rho_{d}}{d a}= 3 \tilde{A} \rho_{d,0} \left( 1 + 3 \tilde{A} (1-\alpha) \ln \frac{a}{a_{0}} \right)^{1/(1-\alpha)-1} \frac{1}{a} \,
\end{equation}
shows that for $\tilde{A} > 0$ the dark energy density grows, for $\tilde{A}=0$ the dark energy is constant,
while for $\tilde{A} < 0$ it decreases with the expansion of the universe.

We assume that the universe contains two components, the dark energy with the EOS (\ref{eq:oureos}) and the nonrelativistic matter. The evolution of the universe is determined by the Friedmann equation
\begin{equation}
\label{eq:Fried}
\left( \frac{\dot{a}}{a} \right)^2 + \frac{k}{a^2} = \frac{8 \pi G}{3} (\rho_{d}+\rho_{m}) \, ,
\end{equation}
where dots denote time derivatives and $\rho_{m}$ stands for nonrelativistic matter energy density.

In the analysis of the dynamics of the universe we first focus on the fate of the universe, i.e., its asymptotic expansion.
We assume that the dark energy density increases, remains constant or decreases more slowly than the nonrelativistic matter energy density or the curvature
term $k/a^2$. The time dependence of the scale factor for $\alpha \neq 1/2$ is then given by the expression
\begin{eqnarray}
\label{eq:aodt}
& &\left( 1 + 3 \tilde{A} (1-\alpha) \ln \frac{a_{1}}{a_{0}} \right)^{\frac{1-2\alpha}{2(1-\alpha)}} \nonumber \\
&-& \left( 1 + 3 \tilde{A} (1-\alpha) \ln \frac{a_{2}}{a_{0}} \right)^{\frac{1-2\alpha}{2(1-\alpha)}} \nonumber \\
&=&\frac{3}{2} \tilde{A} (1-2\alpha) \Omega_{d,0}^{1/2} H_{0} (t_{1}-t_{2}) \, .
\end{eqnarray}
Here  $\Omega_{d,0}=\rho_{d,0}/\rho_{c,0}$, where $\rho_{c,0} = 3 H_{0}^2/(8 \pi G)$ is the critical energy density
at the present epoch.

For $\alpha = 1/2$, the expression for the dependence of the scale factor on time becomes
\begin{equation}
\label{eq:aodtspec}
\ln \frac{1 + \frac{3}{2} \tilde{A} \ln \frac{a_{1}}{a_{0}}}
{1 + \frac{3}{2} \tilde{A} \ln \frac{a_{2}}{a_{0}}} = \frac{3}{2} \tilde{A} \Omega_{d,0}^{1/2} H_{0} (t_{1}-t_{2}) \, .
\end{equation}

Once we have the complete set of equations determining the temporal evolution of the universe, we can study the
parametrical dependence of the cosmological solutions.
Equations (\ref{eq:scalour}), (\ref{eq:aodt}), and (\ref{eq:aodtspec}) single out two specific values for the parameter
$\alpha$: $\alpha=1/2$ and $\alpha=1$. These values represent the bordering points of the intervals on the $\alpha$ line with
qualitatively the same cosmological behavior. Therefore, in our analysis of the parameter space of the model,
for each value of the parameter $\tilde{A}$
we consider the following four intervals for the parameter $\alpha$: $\alpha > 1$, $1 > \alpha > 1/2$, $\alpha=1/2$, and $\alpha <
1/2$. The value $\alpha=1$ is not analyzed since it corresponds to the familiar model of dark energy with the constant parameter of
the EOS. We first consider the positive values of the parameter $\tilde{A}$ which correspond to phantom energy.

{\bf $\tilde{A} > 0$, $\alpha > 1$.} Equation (\ref{eq:oureos}) shows that, with these parameter values, the pressure of dark energy
grows much faster compared with phantom energy with the constant parameter of the EOS. On the basis of this observation, one
might expect even more singular behavior than in the usual ``big rip" case. The expression for the scaling of the dark energy
density with the scale factor $a$, more conveniently written as
\begin{equation}
\label{eq:alphagt1rho}
\rho_{d}=\rho_{d,0} \left( 1 - 3 \tilde{A} (\alpha-1) \ln \frac{a}{a_{0}} \right)^{-1/(\alpha-1)} \, ,
\end{equation}
shows that the dark energy density diverges for the finite value of the scale factor
\begin{equation}
\label{eq:alphagt1amax}
a_{max}=a_{0} e^{1/(3 \tilde{A} (\alpha-1))} \, .
\end{equation}
At the same value of the scale factor the dark energy pressure diverges towards negative infinity. Equation (\ref{eq:aodt}) reveals
that this singularity is reached at finite time $t_{max}$. The evolution of the scale factor before the singularity can be expressed
as
\begin{eqnarray}
\label{eq:abeforetmax}
a &=& a_{0} exp \left[ \frac{1}{3 \tilde{A} (\alpha -1)} \left(1 -\right. \right.
\nonumber \\
& &\left. \left.  \left[ \frac{3}{2} \tilde{A} (2\alpha-1) \Omega_{d,0}^{1/2}
(t_{max}-t) \right]^{\frac{2(\alpha-1)}{2\alpha-1}} \right) \right] \, .
\end{eqnarray}
To summarize, the evolution of the universe reaches the singularity in which the dark energy density and the pressure
diverge at the finite time $t_{max}$ and the {\em finite} value of the scale factor $a_{max}$. 

\begin{figure}
\centerline{\resizebox{0.45\textwidth}{!}{\includegraphics{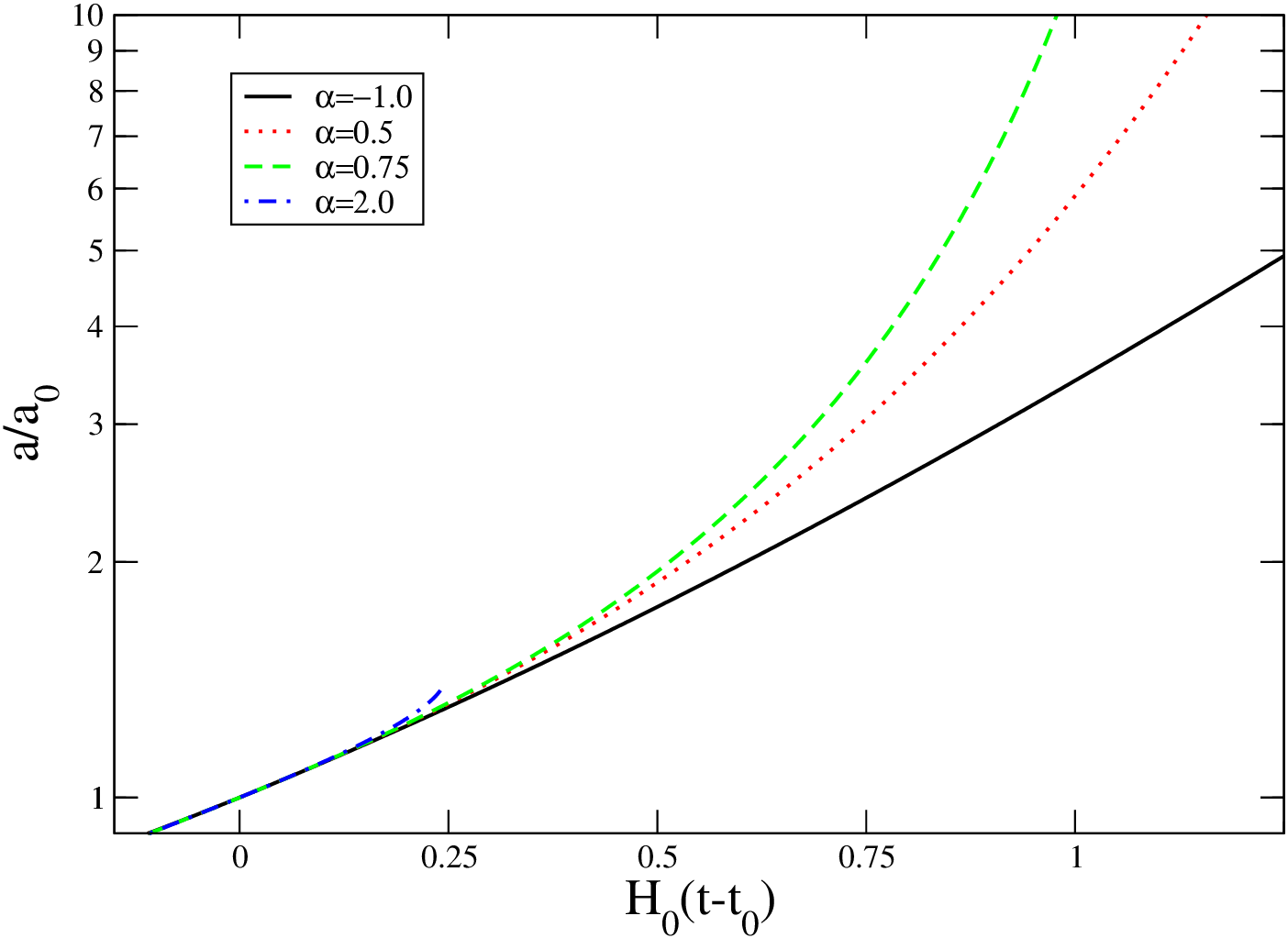}}}
\caption{\label{fig:scalea1}  The time evolution of the scale factor of the universe for
$\Omega_{d,0}=0.7$, $\Omega_{m,0}=0.3$,
$\tilde{A} = 1$ and four typical values of the parameter $\alpha$: $\alpha = -1,0.5,0.75,2$.
For $\alpha=2$, the scale factor reaches the finite value at the singularity while for $\alpha=0.75$, the scale factor of the universe diverges at singularity.}
\end{figure}
\vspace{1cm}

{\bf $\tilde{A} > 0$, $1/2 <\alpha < 1$.} Equation (\ref{eq:scalour}) shows that in this case the model exhibits the minimal value of the scale factor
\begin{equation}
\label{eq:alphagt05amin}
a_{min}=a_{0} e^{-1/(3 \tilde{A} (1-\alpha))} \, .
\end{equation}
Let us denote the instant of time when the description of the evolution of the universe in terms of dark energy only
becomes appropriate
by $t_{*}$ and the corresponding scale factor by $a_{*}$. Equation (\ref{eq:aodt}) acquires the form
\begin{eqnarray}
\label{eq:aodtstar}
& &\left( 1 + 3 \tilde{A} (1-\alpha) \ln \frac{a}{a_{0}} \right) = \nonumber \\
& &\left[ \left( 1 + 3 \tilde{A} (1-\alpha) \ln \frac{a_{*}}{a_{0}} \right)^{-\frac{2\alpha-1}{2(1-\alpha)}} \right. \nonumber \\
& & + \left. -\frac{3}{2} \tilde{A} (2\alpha-1) \Omega_{d,0}^{1/2} H_{0} (t-t_{*}) \right]^{-\frac{2(1-\alpha)}{2\alpha-1}} \, .
\end{eqnarray}
At the instant of time
\begin{equation}
\label{eq:alphagt05trip}
t_{rip}=t_{*}+\frac{2}{3 \tilde{A} (2\alpha -1) \Omega_{d,0}^{1/2} H_{0}} \left(1 +
3 \tilde{A} (1-\alpha) \ln \frac{a_{*}}{a_{0}} \right)^{-\frac{2\alpha-1}{2(1-\alpha)}} \, ,
\end{equation}
the scale factor diverges. The dark energy density and pressure diverge at the same instant. This scenario is qualitatively
equivalent to the ``big rip" scenario present in many phantom energy models.

\begin{figure}
\centerline{\resizebox{0.45\textwidth}{!}{\includegraphics{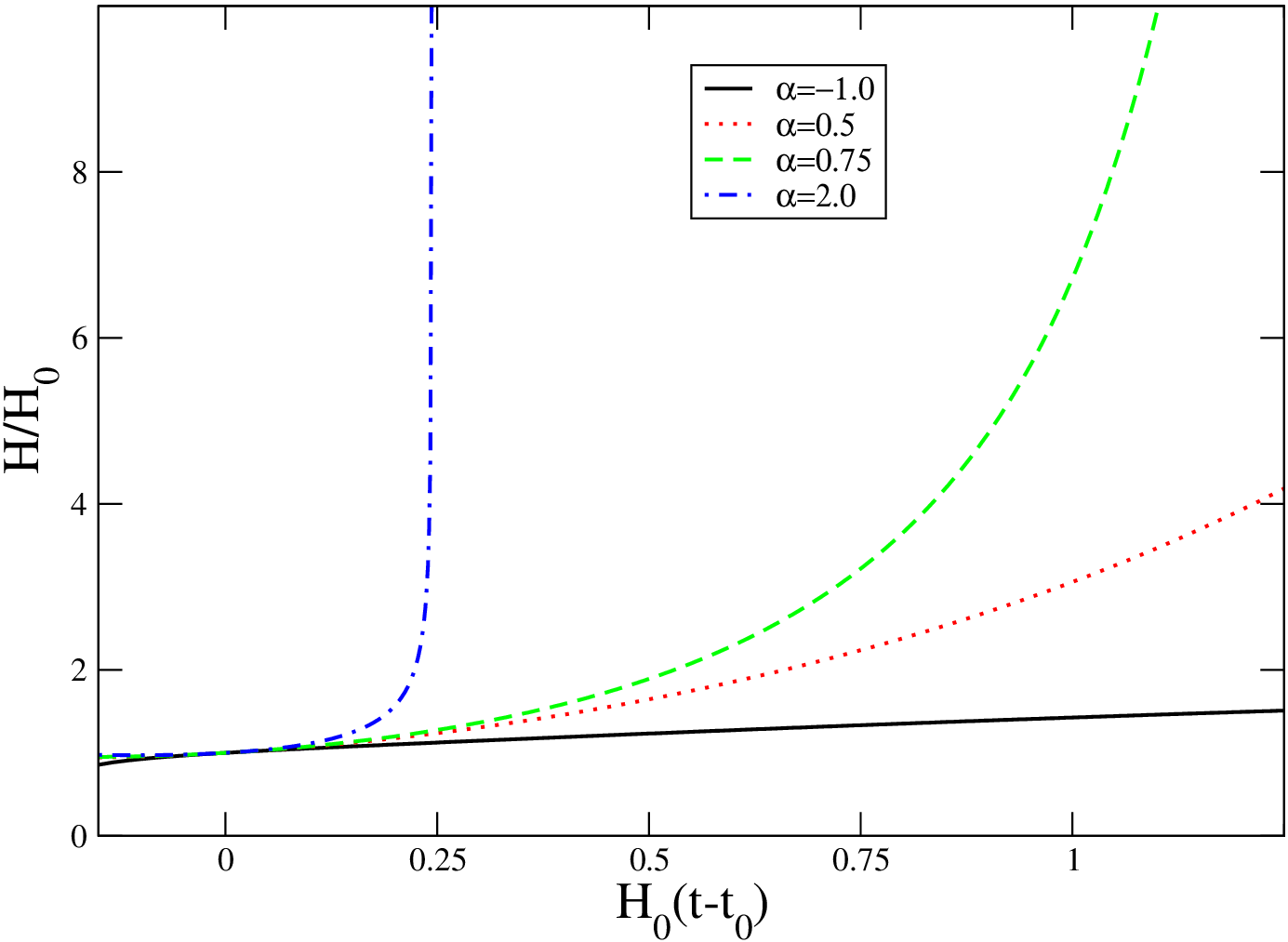}}}
\caption{\label{fig:hubblea1}  The time dependence of the Hubble parameter for $\tilde{A} = 1$,
 $\Omega_{d,0}=0.7$, $\Omega_{m,0}=0.3$,
 and $\alpha = -1,0.5,0.75,2$. For $\alpha=0.75$ and $\alpha = 2$ there is the singularity of the Hubble parameter appearing at finite time.}
\end{figure}
\vspace{1cm}

{\bf $\tilde{A} > 0$, $\alpha = 1/2$.} Equation (\ref{eq:aodtspec}) gives the law of evolution
\begin{eqnarray}
a &=& a_{0} exp \left[ \frac{2}{3\tilde{A}} \left[ \left(1+\frac{3}{2} \tilde{A}
\ln \frac{a_{*}}{a_{0}} \right) \times \right. \right. \nonumber \\
 & & \left. \left. exp \left( \frac{3}{2} \tilde{A}
\Omega_{d,0}^{1/2} H_{0} (t-t_{*}) \right) -1 \right] \right] \, .
\end{eqnarray}
Here the quantities $t_{*}$ and $a_{*}$ have the same interpretation as in the preceding case.
The model exhibits the minimal value of the scale factor $a_{min}=a_{0} exp (-2/(3 \tilde{A}))$. The future expansion of the universe
has no singularities and continues infinitely.

{\bf $\tilde{A} > 0$, $\alpha < 1/2$.} In this case, the model also has a minimal value of the scale factor. The future evolution
of the universe has no singularities and is given by the expression
\begin{eqnarray}
\label{eq:alphalt05aodt}
a&=&a_{0} exp \left[ \frac{1}{3\tilde{A} (1-\alpha)} \left(  \left[
\left( 1 + 3 \tilde{A} (1-\alpha) \ln \frac{a_{*}}{a_{0}} \right)^{\frac{1-2\alpha}{2(1-\alpha)}} \right. \right. \right. \nonumber \\
&+& \left. \left. \left.
\frac{3}{2} \tilde{A} (1-2\alpha) \Omega_{d,0}^{1/2} H_{0} (t-t_{*}) \right]^{\frac{2(1-\alpha)}{1-2\alpha}} -1 \right) \right]
\, .
\end{eqnarray}

\begin{figure}
\centerline{\resizebox{0.45\textwidth}{!}{\includegraphics{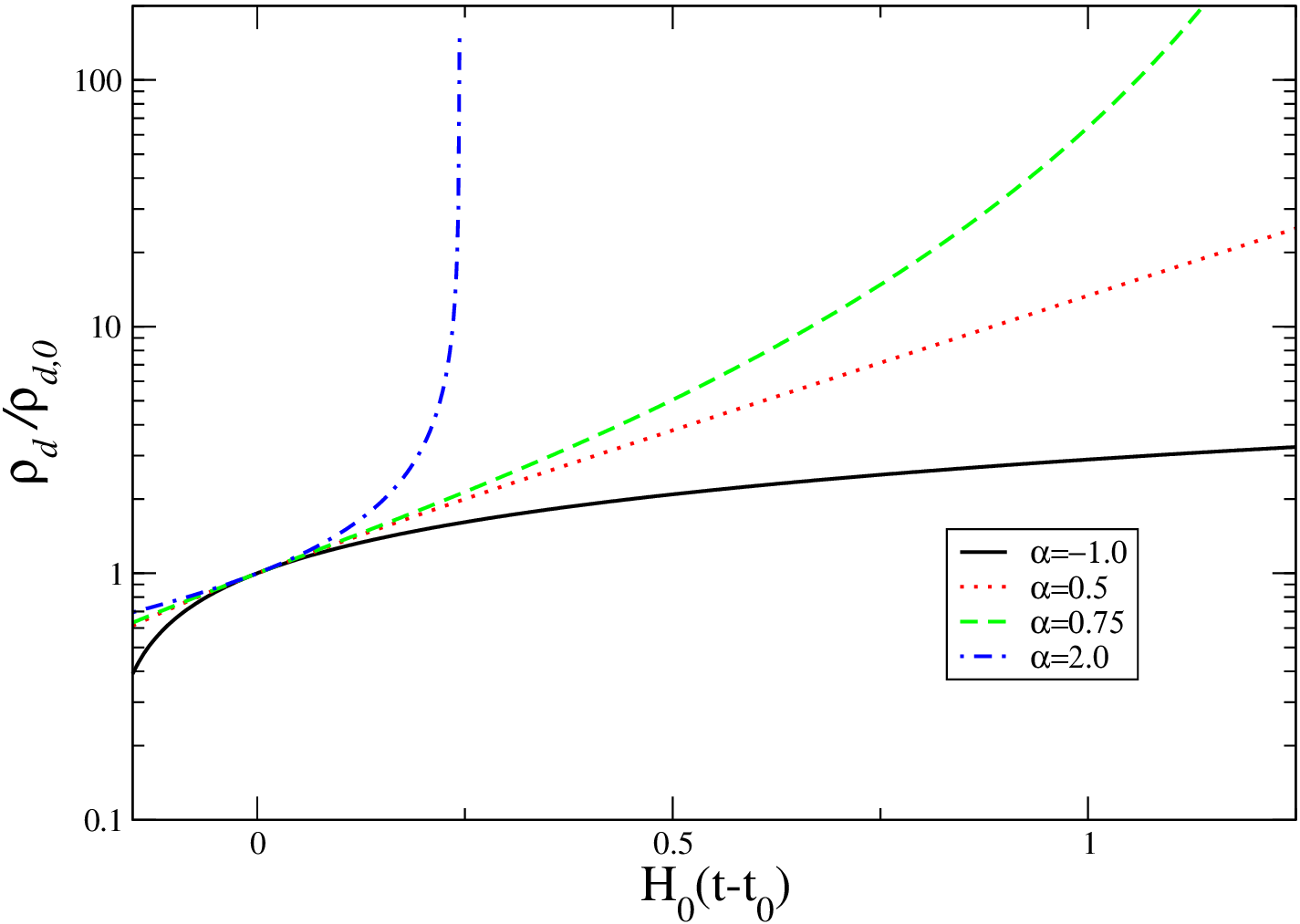}}}
\caption{\label{fig:densitya1}  The time evolution of the dark energy density for $\tilde{A} =1$, $\Omega_{d,0}=0.7$, $\Omega_{m,0}=0.3$,
and the typical values $\alpha= -1,0.5,0.75,2$. For $\alpha=0.75$ and $\alpha = 2$, there is the onset of singularity of the dark energy density at finite time.}
\end{figure}

\begin{figure}
\centerline{\resizebox{0.45\textwidth}{!}{\includegraphics{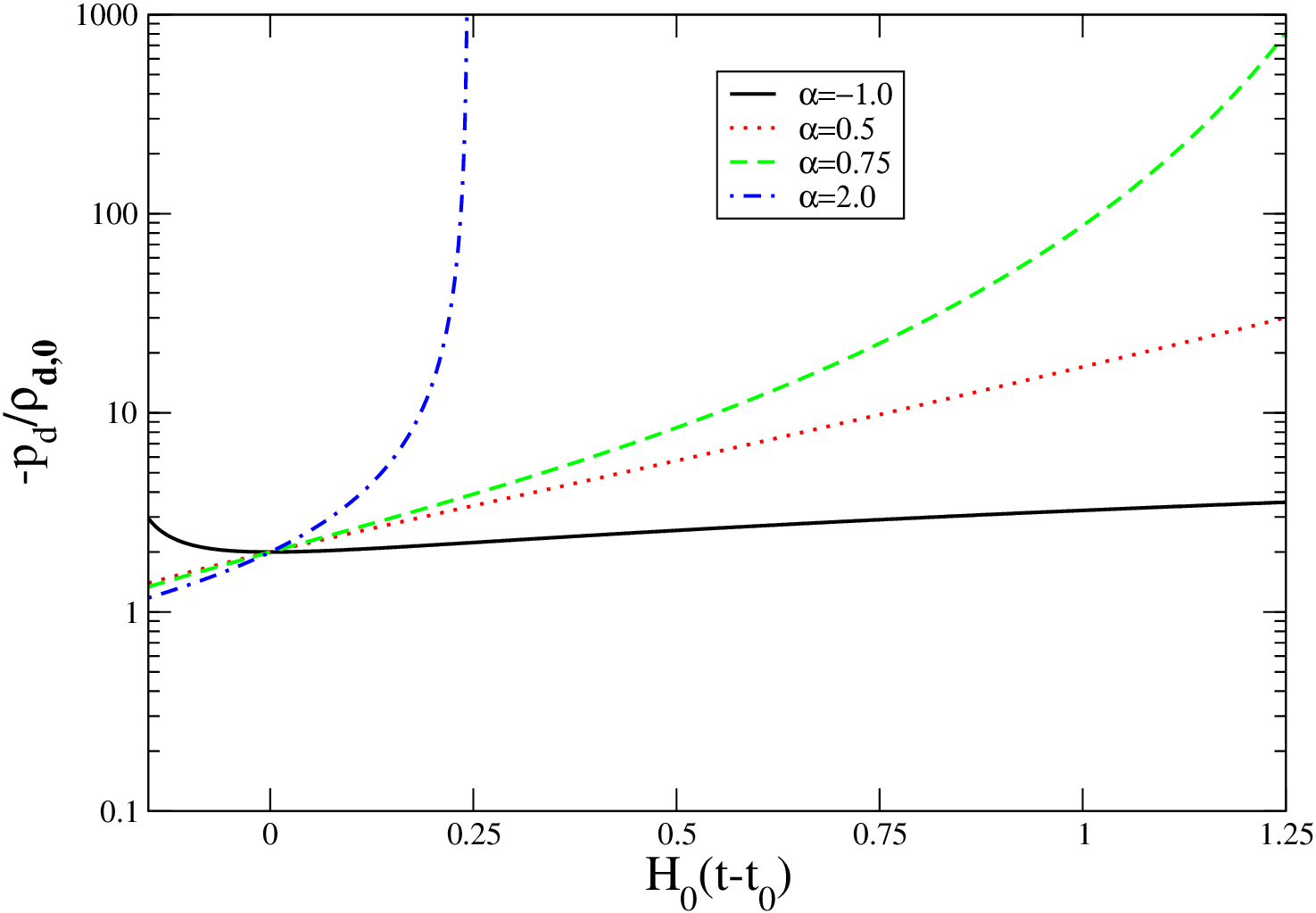}}}
\caption{\label{fig:pressurea1}  The time dependence of the dark energy pressure for $\tilde{A} = 1$, $\Omega_{d,0}=0.7$, $\Omega_{m,0}=0.3$, and the typical values $\alpha= -1,0.5,0.75,2$. For $\alpha=0.75$ and $\alpha = 2$, the dark energy pressure diverges to $-\infty$. }
\end{figure}

The results on the asymptotic expansion of the universe, obtained in an analytically tractable form, can be further supported and completed by numerical calculations.
The full time dependence of the universe expansion obtained numerically is depicted in Figs
\ref{fig:scalea1}, \ref{fig:hubblea1}, \ref{fig:densitya1}, and \ref{fig:pressurea1}.

The analysis presented so far has been concentrated on the models with positive $\tilde{A}$ which are of phantom character. Next we focus on the models with negative $\tilde{A}$.
In these models the character of dark energy is nonphantom.

\vspace{1cm}

The accelerated expansion of the universe at the present epoch implies the negativity of the dark energy pressure at present. This requirement was automatically satisfied for the models with the phantom character. For models with $\tilde{A} < 0$, this requirement produces nontrivial constraints on the parameter $\tilde{A}$. Namely, if we require the negativity of the dark energy pressure at the pressent epoch, the parameter  $\tilde{A}$ must obey the constraint $\tilde{A} > -1$. If we demand the negativity of the dark energy pressure at some $z > 0$, an analogous condition for $\tilde{A}$ can be obtained in a straightforward manner and it is generally $\alpha$ dependent. The behavior of nonphantom dark energy models is further classified with respect to the parameter $\alpha$.

{\bf $\tilde{A} < 0$, $\alpha > 1$.} The scaling of the dark energy density with the scale factor
\begin{equation}
\label{eq:alphagt1Aless}
\rho_{d}=\rho_{d,0} \left( 1 + 3 \tilde{A} (1-\alpha) \ln \frac{a}{a_{0}} \right)^{-1/(\alpha-1)} \, ,
\end{equation}
points to the existence of the minimal value of the scale factor $a_{min}=a_{0} exp(-1/(3 \tilde{A} (1 -\alpha)))$. The energy density $\rho_{d}$ decreases with the expansion of the universe. The term proportional to $\rho^{\alpha}$ becomes less significant with the expansion of the universe and the dark energy EOS asymptotically approaches the cosmological constant EOS. These findings on the scaling properties of the dark energy density are also reflected in the asymptotic evolution of the scale factor:
\begin{eqnarray}
\label{eq:Alessa}
a &=& a_{0} exp \left[ \frac{1}{3\tilde{A} (1-\alpha)} \left(  \left[
\left( 1 + 3 \tilde{A} (1-\alpha) \ln \frac{a_{*}}{a_{0}} \right)^{\frac{1-2\alpha}{2(1-\alpha)}} \right. \right. \right. \nonumber \\
&+& \left. \left. \left.
\frac{3}{2} \tilde{A} (1-2\alpha) \Omega_{d,0}^{1/2} H_{0} (t-t_{*}) \right]^{\frac{2(1-\alpha)}{1-2\alpha}} -1 \right) \right]
\, .
\end{eqnarray}
The future expansion of the universe is unbounded and nonsingular.

\vspace{1cm}

We continue the discussion with some general remarks for  all cases with $\alpha < 1$
since they share the same type of behavior of the dark energy density. Namely, the dark energy density (\ref{eq:scalour}) decreases and at the scale factor
\begin{equation}
\label{eq:anull}
a_{NULL} = a_{0} e^{-1/(3 \tilde{A} (1-\alpha))} \,
\end{equation}
the dark energy density vanishes. For these parameter values, the expressions for the asymptotic expansion are no longer applicable since dark energy does not dominate the future expansion of the universe.
From the dark energy EOS (\ref{eq:oureos}) we can see that at $a_{NULL}$ the dark energy pressure vanishes for $0 < \alpha < 1$, acquires the finite value $ p_{d} = -\tilde{A}$ for $\alpha=0$ and diverges for $\alpha < 0$. Let us describe the dynamics of the universe in more detail for different intervals of the parameter $\alpha$.

{\bf $\tilde{A} < 0$, $1/2 < \alpha < 1$}. Let us discuss the future behavior of the universe for a simplified cosmological model with a single component which is dark energy with the EOS (\ref{eq:oureos}). In this setting, equations (\ref{eq:aodt}) and (\ref{eq:aodtspec}) become exact for the flat universe. Although the value $a_{NULL}$ is finite, it is reached only at infinite time, as can be seen from (\ref{eq:aodt}). Therefore, in the model with the dark energy component only, there is no finite time singularity. The quantities $\rho_{d}$ and $p_{d}$ asymptotically go to $0$.

{\bf $\tilde{A} < 0$, $\alpha = 1/2$}. In the model with the dark energy component only, introduced in the preceding paragraph, equation (\ref{eq:aodtspec}) governs the expansion of the universe. The expansion reaches $a_{NULL}$ at infinite time and there is no finite time singularity. Both $\rho_{d}$ and $p_{d}$ asymptotically vanish.

{\bf $\tilde{A} < 0$, $ 0 < \alpha < 1/2$}. In the model with the dark energy component only, the universe reaches the finite scale value $a_{NULL}$ at finite time
$t_{NULL}=-2/(3 \tilde{A} (1-2\alpha) \Omega_{d}^{1/2} H_{0}) $, as follows from (\ref{eq:aodt}). At $t_{NULL}$ the quantities $\rho_{d}$ and $p_{d}$ vanish. The expansion of the universe stops with $\dot{a}(t_{NULL}) = 0$ and $\ddot{a}(t_{NULL})=0$. There is no singularity at the finite time $t_{NULL}$.

\begin{figure}[floatfix]
\centerline{\resizebox{0.45\textwidth}{!}{\includegraphics{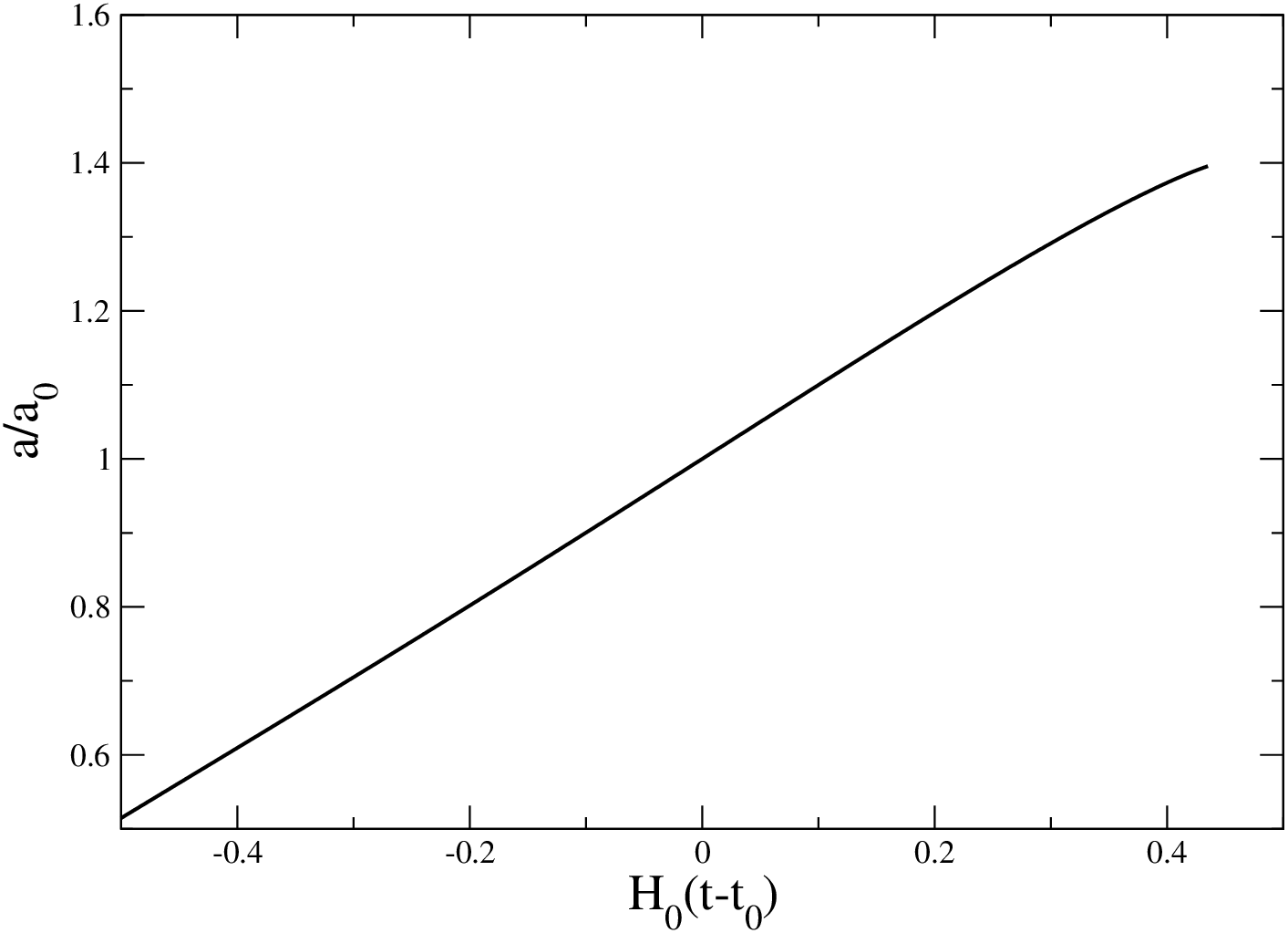}}}
\caption{\label{fig:scaleBar}  The time evolution of the scale factor of the universe for
 $\Omega_{d,0}=0.7$, $\Omega_{m,0}=0.3$,
 $\tilde{A}=-0.5$ and $\alpha=-1$. The scale factor of the universe is finite at the onset of singularity.}
\end{figure}
\vspace{1cm}

In the interval $0 < \alpha < 1$, the scaling of $\rho_{d}$ with $a$ is generally not well defined by (\ref{eq:scalour}) for $a > a_{NULL}$. However, for countably many $\alpha$ values the expression (\ref{eq:scalour}) can be extended beyond $a_{NULL}$. Namely, for values $\alpha=1-1/(2n)$, where $n = 1,2,\dots$, the dark energy density is well defined for all scale factor values.
If we allow the possibility of negative values for the dark energy density, for $\alpha = 1-1/(2n+1)$, where $n=0,1,2,\dots$, the dark energy density is well defined for all scale values as well. However, for negative values of $\rho_{d}$, the dark energy pressure is not well defined.

The four preceding paragraphs discuss the evolution of the universe in a model containing only the dark energy component with the EOS (\ref{eq:oureos}). A realistic model with both nonrelativistic and dark energy components can be easily numerically handled for scale factors up to $a_{NULL}$. Furthermore, the numerical analyses show that $a_{NULL}$ is reached at finite time for parameter values $0 < \alpha < 1$. At $a_{NULL}$ the quantities $\rho_{d}$ and $p_{d}$ vanish, but $\rho_{m}$ and $H$ do not. At $a_{NULL}$ there is no singularity, which is an expected feature. Namely, in the situation where the dark energy component does not cause any singularity, it is highly improbable that the addition of the well-behaved nonrelativistic component would induce singular behavior. The situation at $a_{NULL}$ implies that the expansion should continue beyond $a_{NULL}$. However, at $a>a_{NULL}$ the scaling of $\rho_{d}$ with $a$ is not well defined (\ref{eq:scalour}). It is important to notice that the scaling equation (\ref{eq:scalgen}) is also satisfied with $\rho_{d}=0$ (and $p_{d}=0$). Therefore, a possible explanation of the expansion at $a>a_{NULL}$ is that once the dark energy density reaches $0$ at $a_{NULL}$, it remains in the state with $\rho_{d}=0$ and $p_{d}=0$. The expansion at $a>a_{NULL}$ is entirely governed by the nonrelativistic matter and the dark energy is gravitationally ineffective. This interpretation yields a cosmological model in which the expansion at $a>a_{NULL}$ is decelerated, and (for a flat universe) the expansion continues to infinity. Furthermore, during the entire expansion, there are no singularities in $\dot{a}$ and $\ddot{a}$.

\begin{figure}
\centerline{\resizebox{0.45\textwidth}{!}{\includegraphics{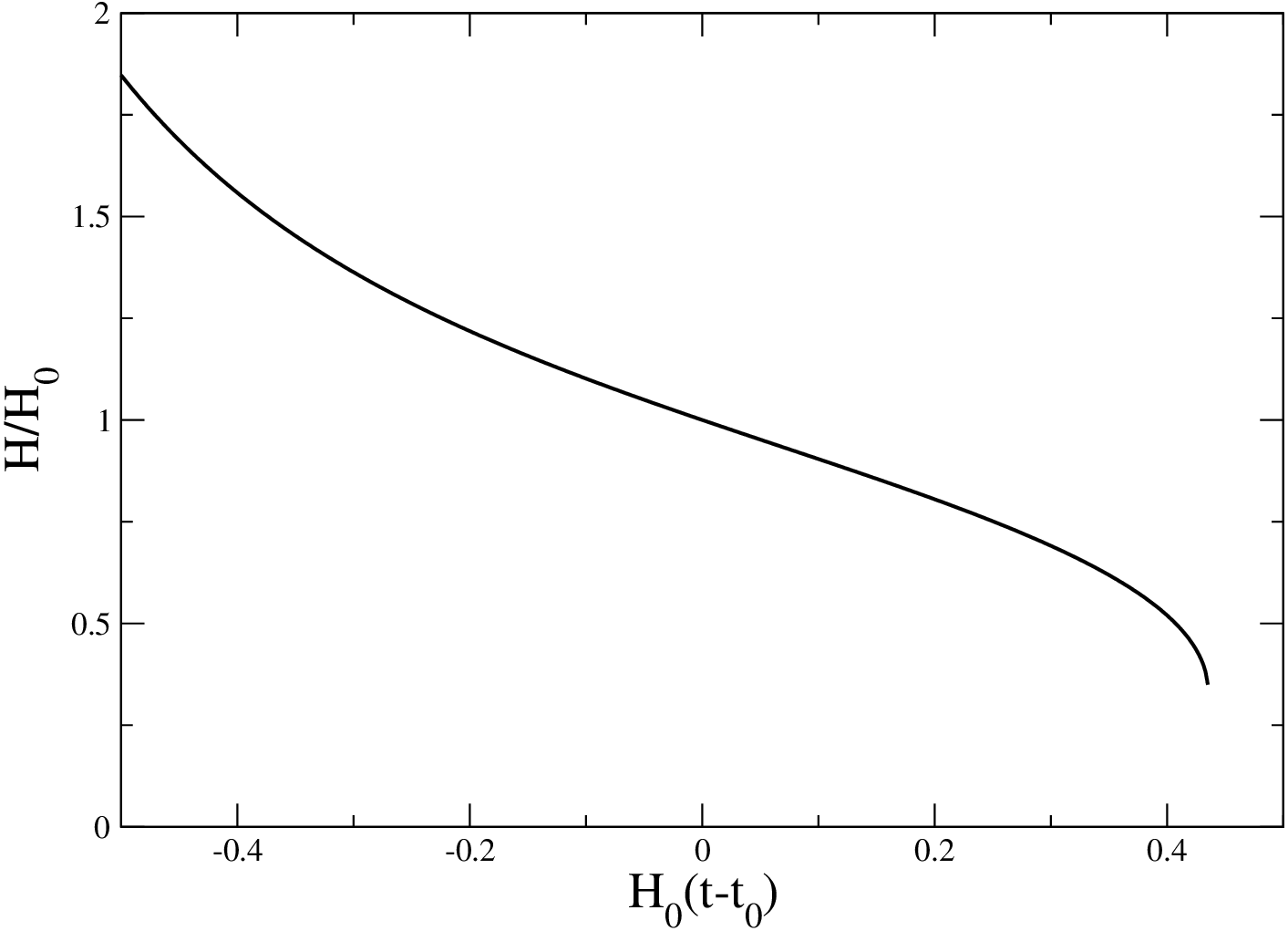}}}
\caption{\label{fig:hubbleBar} The dependence of the Hubble parameter on time for
$\Omega_{d,0}=0.7$, $\Omega_{m,0}=0.3$,
$\tilde{A}=-0.5$ and $\alpha=-1$. The Hubble parameter is finite at the onset of
singularity.}
\end{figure}

\begin{figure}
\centerline{\resizebox{0.45\textwidth}{!}{\includegraphics{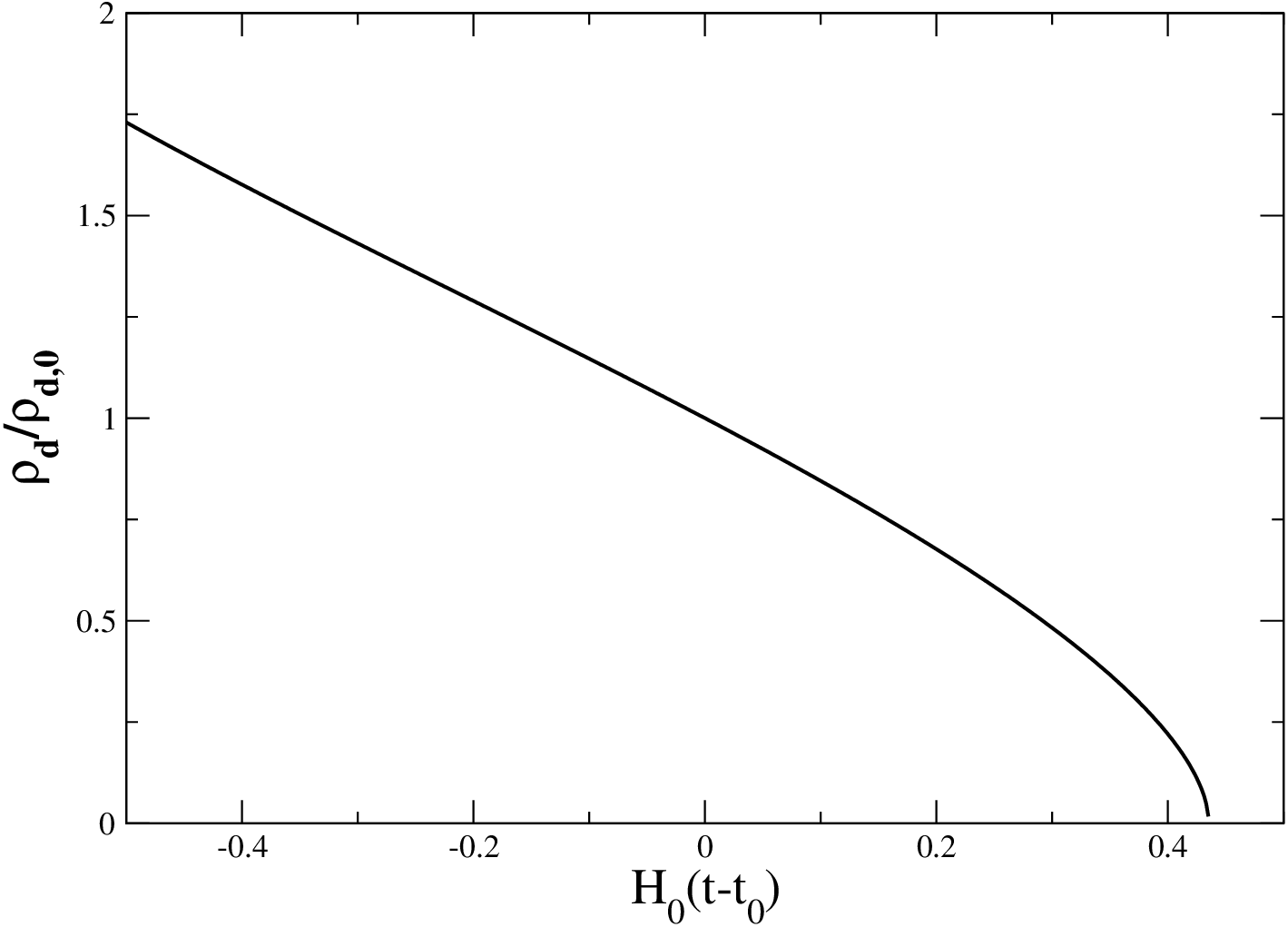}}}
\caption{\label{fig:densityBar} The dark energy density dependence on time for
$\Omega_{d,0}=0.7$, $\Omega_{m,0}=0.3$,
$\tilde{A}=-0.5$ and $\alpha=-1$. The dark energy density vanishes at the singularity. }
\end{figure}

\begin{figure}
\centerline{\resizebox{0.45\textwidth}{!}{\includegraphics{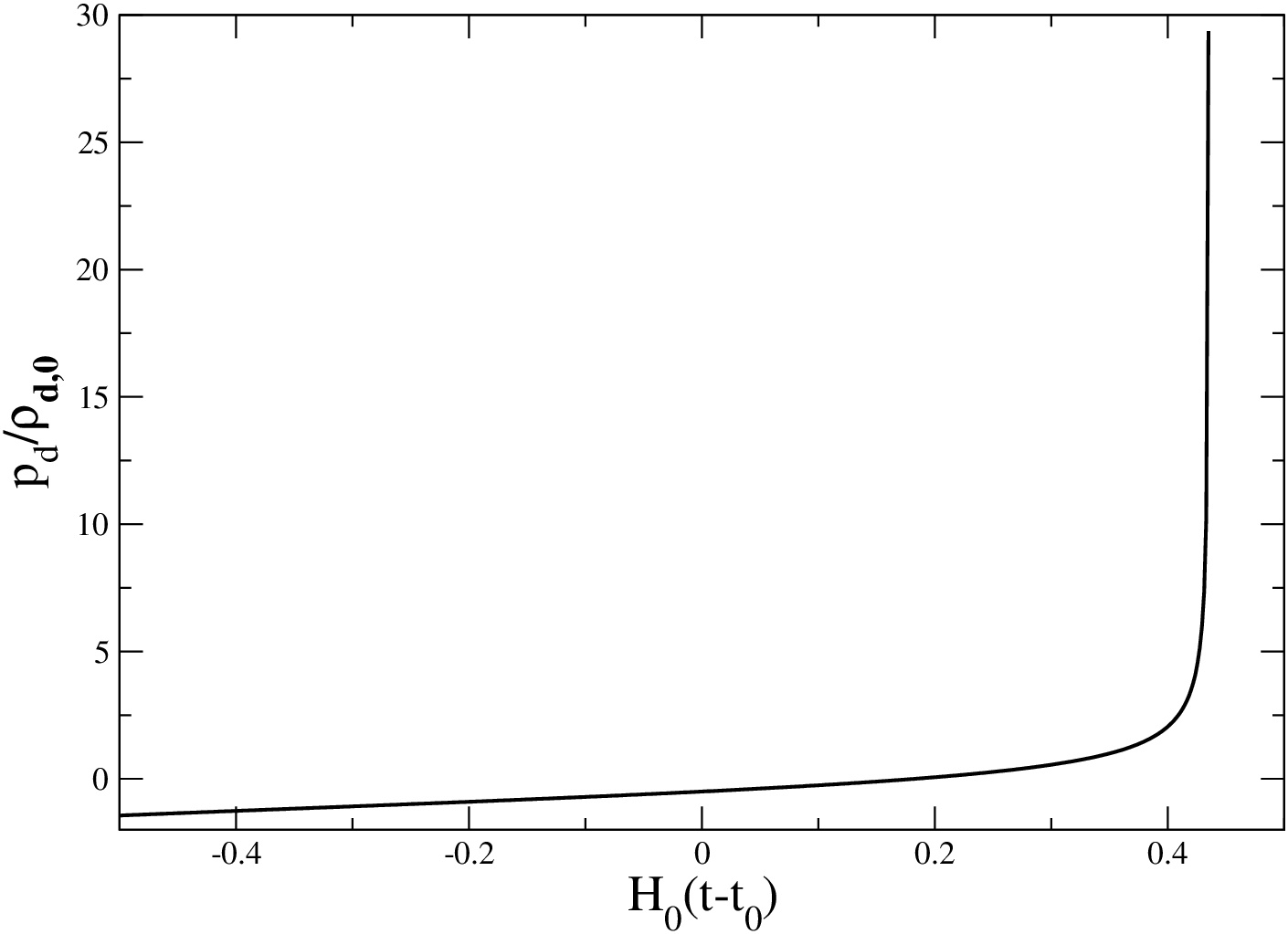}}}
\caption{\label{fig:pressureBar} The dark energy pressure dependence on time for
 $\Omega_{d,0}=0.7$, $\Omega_{m,0}=0.3$,
 $\tilde{A}=-0.5$ and $\alpha=-1$. The dark energy pressure diverges towards $+\infty$. }
\end{figure}

{\bf $\tilde{A} < 0$, $\alpha < 0$.} The dark energy density tends to $\rho_{d}=0$ as the scale factor approaches $a_{NULL}$. At the same time the pressure diverges, i.e., $p_{d} \rightarrow \infty$. In the model which contains only the dark energy component, this singularity is reached at a finite moment $t_{NULL}=-2/(3 \tilde{A} (1-2\alpha) \Omega_{d}^{1/2} H_{0}) $. The numerical calculations for the more realistic model with both nonrelativistic matter and dark energy components also show that the singularity is reached at finite time. The time dependences of relevant cosmological quantities are displayed in Figs
\ref{fig:scaleBar}, \ref{fig:hubbleBar}, \ref{fig:densityBar}, and \ref{fig:pressureBar}. The characteristics of this singularity are as follows: the singularity is encountered at finite time; at the singularity the scale factor is finite, Fig. \ref{fig:scaleBar}; the quantities $\dot{a}$ and $H$ are finite, Fig. \ref{fig:hubbleBar}, and so are the energy densities $\rho_{d}$ (Fig. \ref{fig:densityBar}) and $\rho_{tot}=\rho_{m}+\rho_{d}$; the acceleration and the dark energy pressure (which equals the total pressure) diverge (Fig. \ref{fig:pressureBar})
\begin{equation}
\label{eq:Barrsing}
\ddot{a} \rightarrow -\infty \, , \;\;\;\;\; p_{d} \rightarrow + \infty \, .
\end{equation}
The characteristics of this singularity match those of the sudden future singularity introduced by
Barrow \cite{Barrow,sudden}. Therefore, the cosmology with the dark energy which has the
EOS (\ref{eq:oureos}) for $\tilde{A} < 0$ and $\alpha < 0$ leads to the finite sudden singularity which
satisfies the strong energy condition. This example shows that a relatively simple dark energy EOS may lead to
the type of singularity proposed by Barrow. Some other types of cosmological singularities characterized
by an infinite deceleration have recently been introduced \cite{shtgor}.

From the overview of the different types of the expansion of the universe, we can see that in various parameter regimes the dark energy model (\ref{eq:oureos}) leads to various types of singularities or asymptotic behaviors. The wealth of phenomena embodied in one model makes it an attractive object of theoretical study. At this moment we would like to emphasize that the model has some advantages beyond theoretical considerations. Namely, dark energy with the EOS (\ref{eq:oureos}) is very suitable for the analysis of cosmological data. The dark energy EOS is centered around the cosmological constant EOS and the parameters $\tilde{A}$ and $\alpha$ measure the deviation from the cosmological constant EOS. The sign of the parameter $\tilde{A}$ determines whether dark energy is of the phantom or the quintessence character. Furthermore, the parameter $\tilde{A}$ determines the amplitude of deviation from the cosmological constant EOS, while the parameter $\alpha$ determines the speed of variation of this deviation.
It could be said that the parameters $\tilde{A}$ and $\alpha$ measure the deviation from the
$\mathrm{\Lambda CDM}$ cosmology.
Different intervals of the parameter $\alpha$, for the fixed $\tilde{A}$, imply the qualitatively different destiny of the universe. The constraint from the observational data to any of the intervals for the parameter $\alpha$ could qualitatively determine the fate of the universe and eliminate some types of singularities or asymptotic behaviors. Finally, this dark energy model provides a natural framework for the time-varying parameter of the EOS. All these characteristics put forward the dark energy model (\ref{eq:oureos}) as a promising candidate for the analysis of cosmological data.

\vspace{1cm}

The analysis presented so far is based upon the expression (\ref{eq:oureos}) as an {\em exact} EOS. However, the usefulness
of the results of this analysis surpasses the model itself. It would certainly be of interest to investigate the models obtained by adding
new terms into the dark energy EOS (\ref{eq:oureos}). This would lead to a general dark energy EOS. Even in a general dark energy model
the results of this paper are still useful, since the EOS (\ref{eq:oureos}) can in this case be considered as an approximation of the more general
EOS around the present epoch of the evolution of the universe. In this case, the solutions obtained above describe the expansion of the universe around
the present epoch well, whereas the description of the singularities obtained in the analysis of (\ref{eq:oureos}) is generally no longer
applicable. The analysis of the dark energy model presented in this paper reveals that even a relatively simple dark energy EOS may comprise very different
types of singularities. It is reasonable to expect that for a model with a more general EOS we may expect even much richer abundance of phenomena. A very
concrete contribution of the EOS (\ref{eq:oureos}) is the demonstration which EOS may lead to the type of singularity at finite time and finite scale factor
and to an opposite type of the sudden future singularity \cite{Barrow}. Another important conclusion is that both of these singularities follow from the same dark
energy EOS, but in different parameter regimes.

\vspace{1cm}

The considerations of the model with the EOS (\ref{eq:oureos}) presented so far have been devoted to the future
evolution of the universe. Let us briefly discuss the consequences of the dark energy model considered in
this paper for the past evolution of the universe and the constraints on the model parameters which follow
from the comparison againts the data obtained from various cosmological observations. For $\tilde{A}>0$ and
$\alpha > 1$ the dark energy density grows with the expansion of the universe, as does the absolute value of 
the dark energy pressure. Therefore, at large redshifts
the dark energy density and pressure become negligible compared to the energy densities of other components 
of the universe and negligibly affect the physics of the early universe (such as primordial nucleosynthesis).
For the case $\tilde{A}>0$ and $\alpha<1$ there is the minimal value of the scale factor $a_{min}$
(\ref{eq:alphagt05amin}) at which the dark energy vanisehes and is not well defined for smaller scale factors.
Furthermore, for $\alpha < 0$, the absolute value of the dark energy pressure becomes very large and diverges at
$a_{min}$. In order to avoid significant impact on the early universe dynamics, the minimal scale factor 
of the model needs to be very small. From (\ref{eq:alphagt05amin}) follows that $\tilde{A} (\alpha-1)$ must
be a very small number. In the case $\tilde{A}<0$ and $\alpha>1$ there is the minimal value of the scale factor
$a_{min}$ again. In this case, however, the dark energy density diverges. Clearly, $a_{min}$ has to be very small 
in order to maintain the consistency with the results of the early universe physics. This again restricts the 
product $\tilde{A} (\alpha-1)$ to be a sufficiently small number.
For the case $\tilde{A}<0$ and $\alpha < 1$,
the dark energy density diverges as the scale factor goes to 0 and since its dependence on the scale factor is
of the logarythmic type, no severe constraints are expected for finite scale factor values. Finally, it is 
important to say that these considerations on the past behavior of the universe are made under the 
assumption that the the dark energy EOS is valid even for the smallest values of the scale factor. If one 
considers (\ref{eq:oureos}) only as a late time effective dark energy EOS, the conditions on parameters $\alpha$
and $\tilde{A}$ may be much less stringent.

\vspace{1cm}

The treatment of this paper is entirely classical. In the vicinity of singularities the quantum effects come 
into play, i.e., the universe enters a new quantum gravity era \cite{Odin}. It has been shown \cite{Odin} 
that quantum effects may reduce or even completely remove the singular behavior at classical singularities.
In that respect, the study of the interplay of classical and quantum effects is needed to obtain a complete
physical picture of the dynamics near the classical singularity.

In conclusion, the dark energy model with the EOS (\ref{eq:oureos}) exhibits a number of interesting phenomena in different parameter regimes. The model comprises several types of singularities appearing at some finite time after the present epoch of the evolution of the universe. For $\tilde{A} > 0$ and $\alpha > 1$, the universe encounters the singularity at  finite time and {\em finite} value of the scale factor, where the dark energy density and pressure diverge. For $\tilde{A} > 0$ and $1/2 < \alpha < 1$, at singularity, which is reached
at finite time, the scale factor and the dark energy density and pressure diverge. This singularity is in these respects equivalent to the ``big rip" singularity found for phantom energy models with the constant parameter of EOS. For $\tilde{A} < 0$ and $\alpha < 0$, the singularity is reached at finite time and finite value of the scale factor. At the singularity the dark energy density vanishes and the total energy density of the universe is finite. The dark energy pressure diverges to $+\infty$ at the singularity. This form of singular behavior is of the type of sudden future singularity introduced by Barrow \cite{Barrow}. The model comprises a number of singularity types with very different properties. Furthermore, it provides a relatively simple framework for the concrete realization of the sudden future singularities such as those introduced by Barrow. In the rest of parameter space, the expansion of the universe is nonsingular with, however, different asymptotic behavior. The model itself is a useful testing ground for the study of singularities in
dark energy cosmologies. The dark energy with the EOS (\ref{eq:oureos}) has an additional advantage that the restrictions on the parameters $\tilde{A}$ and $\alpha$ obtained by  comparison with the observational data may qualitatively select the destiny of the universe. The model also provides a natural framework for the study of dark energy with a time-dependent parameter of the EOS. The features of the dark energy model (\ref{eq:oureos}) imply that for some more general dark energy EOS one might expect an even larger abundance of interesting phenomena and singularities discussed in the literature may appear for the relatively simple dark energy EOS. The comparison with the observational data will hopefully determine which destiny of our universe is the most reasonable to expect.


{\bf Note added.} In a recent paper \cite{odingener}, appearing shortly after our paper, a general analysis
of the types of singularities in phantom energy models was made. The singularity at the finite time and the
finite value of the scale factor, first introduced in our paper, was classified as the type III singularity in
\cite{odingener}. Furthermore, using the integrated conformal anomaly,  an analysis of quantum 
effects near singularity, was performed in \cite{odingener}. It was shown that for the singularity of the 
type III appearing for the model (\ref{eq:oureos}) the behaviour near the classical singularity becomes less 
singular. In particular, the quantum effects near the type III singularity may completely remove the 
singularity in the behavior of the Hubble parameter $H$. The reduction of the singular behavior owing to the 
quantum effects was also found for other types of singularities in phantom energy models.

{\bf Acknowledgements.} The autor would like to thank N. Bili\'{c}, B. Guberina, R. Horvat, and 
J. Sol\`{a} for useful comments and communications. This work was supported by the Ministry of Science, 
Education and Sport of the Republic Of Croatia under the contract No. 0098002.



\end{document}